# Multi-epoch Radio Source Structure Analysis of 11 Calibrators at 2.3 and 8.4 GHz in the South

**Sanway Chatterjee**
Rasulpur, Memari, Purba Bardhaman
West Bengal, PIN 713151, India
iamsanway@gmail.com

**Sayan Basu**
Wits Centre for Astrophysics, School of Physics
University of the Witwatersrand
Private Bag 3, 2050, Johannesburg, South Africa

**Daniel MacMillan**
NVI, Inc., NASA Goddard Space Flight Center
Code 61A, Greenbelt, MD, USA

January 12, 2023

## Abstract

We present the source structure analysis of 11 calibrator sources below $-40°$ south at 2.3 (S-band) and 8.4 GHz (X-band). We used multi-epoch very long baseline interferometry source maps available in the radio fundamental catalog to analyse jet-structure variability and also used fluxes from the Goddard Space Flight Center database to see whether these two complement each other or not. Also, total fluxes from the maps were plotted with the fluxes from the database. The S/X-band light curve analysis provides a more clear picture of the structural variability at the S/X-band also indicates the possibility of the "core-shift" phenomenon. We found jet-like structures in the majority of the sources in the sample.

*Keywords* :

Radio astronomy – quasars – interferometry – source structure

## 1 Introduction

Extragalactic radio sources used in radio catalogs such as the International Celestial Reference Frame (ICRF; Ma et al., 1998) or Radio Fundamental Catalog[1] (RFC), are generally active galactic nuclei (AGN) with an active supermassive black hole at the centre. Extragalactic radio sources, in general, exhibit time- and frequency-dependent source structures. The structure of these sources varies with time, it is therefore important to model their structure at multiple epochs in order to define a time-dependent source model. Quasars, being the brightest in the AGN subclasses and also being located at far distances from the earth (e.g., the most distant quasar yet identified is J0313-1806 at redshift z = 7.642; Wang et al., 2021), show almost no proper motion on the sky. Therefore, being the brightest and appearing as a point-like source in the sky with no proper motion, they are considered to be good candidates as calibrators.

High angular resolution maps from Very Long Baseline Interferometry (VLBI) observations provide us an opportunity to detect source structures at milliarcsecond (mas) scales. The VLBI positions of quasars are also used to define and maintain the accuracy of the ICRF. The present catalog (ICRF-3; Charlot et al., 2020) currently contains VLBI positions of 4,536 radio sources (mainly quasars). Among these observed sources, 2,615 (57.65%) are in the Northern Hemisphere and 1,921 (42.35%) are in the Southern Hemisphere. The very long baseline array calibrator surveys (VCS; Beasley et al., 2002, Fomalont et al., 2003, Petrov et al., 2005, Petrov et al., 2006, Kovalev et al., 2007, Petrov et al., 2008, Petrov, 2016) have been used to increase the number of calibrators in the Northern Hemisphere. However, the long baseline array calibrator survey (LCS; Petrov et al., 2011, Petrov et al., 2019), which is dedicated to increase the number of sources as well as to study the VLBI positions, contributed significantly in the south to increase the number of calibrators. Sources selected from the ICRF, are also being selected from the RFC that contains VLBI positions of a total of 20,250 radio sources, where 11,462 sources (56.60%) are in the Northern Hemisphere and 8,788 sources (43.40%) are in the Southern Hemisphere. Despite all these surveys, no dedicated initiatives have been taken yet to study source structure in the Southern Hemisphere routinely.

---

[1] http://astrogeo.org/rfc/

| B1950 Name | J2000 Name | Optical ID | z | RA (hh mm ss) | DEC (deg mm ss) | $S_t(S)$ | $S_t(X)$ | $S_p(S)$ | $S_p(X)$ | Latest epoch |
|---|---|---|---|---|---|---|---|---|---|---|
| 0048-427 | J0051-4226 | QSO | 1.749 | 00 51 09.501827 | -42 26 33.29329 | 0.405 | 0.860 | 0.380 | 0.616 | 2021.01.27 |
| 0104-408 | J0106-4034 | QSO | 0.584 | 01 06 45.107971 | -40 34 19.96031 | 0.884 | 1.154 | 0.868 | 1.097 | 2018.08.10 |
| 0332-403 | J0334-4008 | QSO | 1.445 | 03 34 13.654488 | -40 08 25.39791 | 1.285 | 0.961 | 1.257 | 0.975 | 2021.05.19 |
| 0537-441 | J0538-4405 | QSO | 0.894 | 05 38 50.361557 | -44 05 08.93893 | 2.034 | 1.813 | 1.406 | 1.535 | 2021.05.19 |
| 1104-445 | J1107-4449 | QSO | 1.598 | 11 07 08.694118 | -44 49 07.61837 | 1.603 | 1.179 | 1.360 | 0.675 | 2021.01.27 |
| 1349-439 | J1352-4412 | QSO | 0.050 | 13 52 56.534938 | -44 12 40.38769 | 0.224 | 0.389 | 0.179 | 0.353 | 2021.05.19 |
| 1424-418 | J1427-4206 | QSO | 1.522 | 14 27 56.297561 | -42 06 19.43769 | 0.893 | 1.090 | 0.499 | 1.067 | 2021.01.27 |
| 1451-400 | J1454-4012 | QSO | 1.810 | 14 54 32.912361 | -40 12 32.51452 | 0.367 | 0.541 | 0.085 | 0.377 | 2018.07.31 |
| 2052-474 | J2056-4714 | QSO | 1.489 | 20 56 16.359815 | -47 14 47.62776 | 1.896 | 1.881 | 1.518 | 1.447 | 2017.09.06 |
| 2106-413 | J2109-4110 | QSO | 1.406 | 21 09 33.188592 | -41 10 20.60545 | 0.903 | 0.581 | 0.439 | 0.247 | 2018.05.19 |
| 2333-415 | J2336-4115 | QSO | 1.058 | 23 36 33.985083 | -41 15 21.98402 | 0.468 | 0.676 | 0.305 | 0.530 | 2018.08.10 |

**Table 1:** The physical properties of the selected sources. Right ascension (RA) and declination (Dec) are shown with the most recent positions in the RFC database. Optical identification and redshift (z) are taken from the NASA/IPAC database. The parameter $S_t(S)$ is the total flux in S-band; $S_t(X)$ is the total flux in X-band; $S_p(S)$ is the peak flux in S-band; $S_p(X)$ is the peak flux in X-band.

Being motivated by this problem, we tried to analyze the radio source structure using the available VLBI source maps from the RFC and to complement the analysis of these maps with the fluxes from the Goddard Space Flight Center (GSFC) database. The GSFC database contains the fluxes of the observed sources from all the available VLBI geodetic and astrometric observing sessions (generally 24 hours in duration) from the past several decades. Available multi-epoch VLBI source maps of the selected sources from the RFC have been used to see whether their flux density variability agrees with the flux variability from the light-curves generated using the fluxes available in the GSFC database.

## 2 Observation and Methodology

We have selected sources in the Southern Hemisphere in the declination zone $[-40°, -90°]$ which were observed at 2.3 (S-band) and 8.4 GHz (X-band). Since we are trying to analyze the radio source structure in the calibrator sources at multi-epoch observations, we selected 11 sources which have been observed in more than 10 epochs (Table 1).

To construct light curves and analyse flux variability, we have used the flux densities of the selected sources available in the RFC database. In addition to the RFC database, we used VLBI fluxes of the selected sources from the GSFC database. The database contains S/X-band fluxes of ICRF sources observed in geodetic and astrometric VLBI sessions around the world. Finally, we have used the available multi-epoch VLBI images to see if the flux density variability detected from the light curve appears in the source maps.

## 3 Results and Analysis

One of the characteristics of a calibrator source is its stability in flux density (no or very little flux density variation) in time- and frequency domain. An ideal calibrator appears to be a compact or point-like source over all projected baselines.

In this section, we present results and a detailed analysis of the sources from our sample. For that purpose, we use a metric, flux variability index as well as we constructed light-curves using flux densities available on GSFC database and flux densities obtained from the available VLBI source maps in the RFC database (rfc_2022c). Lastly, we analyse the flux density variability trend (if any) between the flux density from the database and source structures in the VLBI source maps to see whether modelling of source structure can actually be useful to detect structure variability.

### 3.1 Flux Variability Index

The flux variability index is a statistical measure that is indicative of how the series of fluxes of a given source are scattered around the mean flux. To analyse flux density variability in our sample, we used the GSFC database. This metric is used to compare the flux dispersion of different sources. Unlike the standard deviation, which is always to be considered in the context of the mean value, the flux variability index provides a relatively simple tool to compare different flux data. Using the flux data, we calculate the mean flux density, averaged over all epochs in which the sources were observed (Table 2, column 5 and 8). Flux variability index is the ratio of the standard deviation and the mean of the total flux densities. A value of 0.0 indicates no variation over time. Mathematically, the standard formula of the flux variability index is expressed as:

$$F_I = \frac{\sigma}{\bar{S}} \qquad (1)$$

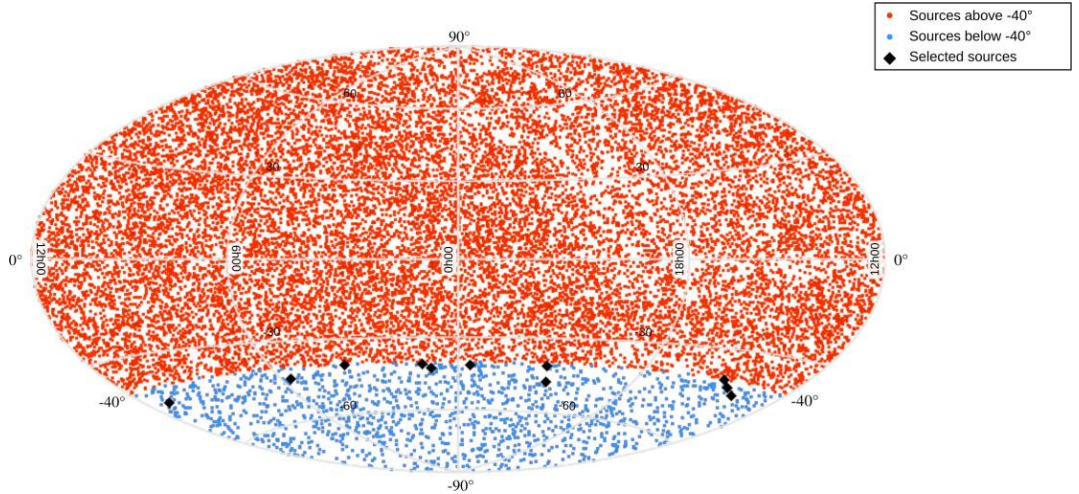

**Figure 1:** Skyplot of all the sources available in the RFC catalog using the Aitoff-hammer projection. The red dots denote the sources above -40 degrees in declination. The blue dots denote the sources below -40 degree which are comparatively much less dense. The black diamonds represent the selected sources used in our analysis.

Where $F_I$ is flux variability index, $\sigma$ and $\bar{S}$ are the standard deviation and the mean of all the flux densities respectively of each source over observing epochs. Using equation (1), we calculate the flux variability indices of all the sources in our sample at S/X-band (Table 2, column 6 and 9).

We present the flux variability index distribution at S-band and X-band in Figure 2 and 3 respectively. At X-band, six sources have a variability index of <0.5. However, at S-band, nine sources have a variability index of <0.5. We also want to see what is the mean and median of the index >0.5. Our results show that selected sources have lower flux density variability at S-band compared to X-band.

### 3.2 Light Curve Analysis

Apart from flux density variability index, we also analysed light curves at S/X-band to understand the trend of flux density variation. The flux densities from the RFC database and GSFC database are used to construct the light-curves. Since the GSFC fluxes are collected on a daily basis from various VLBI observations, a light curve with both the fluxes (from VLBI RFC maps and from the GSFC flux database) can provide us with more information on source structure variability. In our analysis, we have constructed light curves of the selected sources at S/X-band. The total flux (sum of all CLEAN components) from the RFC maps of the selected sources are plotted over epochs of their observation to compare with the variation of the light curves.

Among the selected 11 sources, five sources have flux data for more than 20 years and six sources have flux data for less than 20 years. In most cases, we notice that X-band light curves temporally lead the S-band curves. This time-lag is likely caused by the "core-shift" effect (Shabala et al., 2014). For the source 0537-441, presented here, S- and X-band light curves clearly show that there is a time-lag (Figure 4 and 5).

For example, the source 0537-441 (J0538-4405) has been routinely observed between 2000 and 2020 in 1965 sessions. The source exhibits clear variability in flux-density, and it has multi-epoch VLBI maps to understand the source structure along with the flux-density variation. The source has a flux variability index of 0.6, which indicates variations in fluxes over epochs. Light curve analysis is a useful way to understand these variations. Also, it is useful to quantify the time delay between the flux variations at different frequencies. Figure 4 shows X-band flux density variability over a period of 21 years. In the figure, along with the GSFC fluxes, we plot total fluxes obtained from VLBI maps.

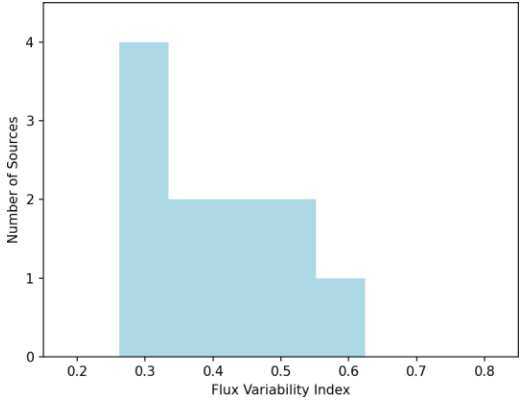
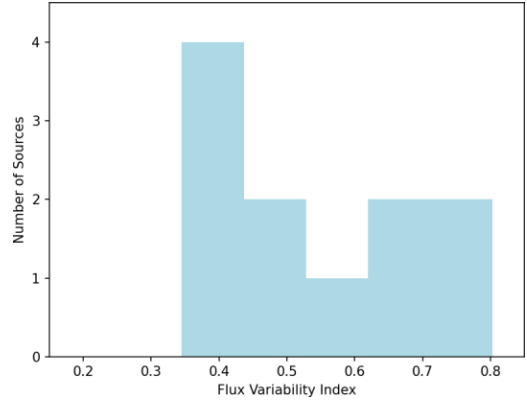

**Figure 2:** Distribution of flux variability index $F_I$ at S-band.  **Figure 3:** Distribution of flux variability index $F_I$ at X-band.

Observing the X-band light-curve indicates that the flux-density varies significantly between MJD 52236 to MJD 52931, where we see flux density changes by a factor of 5.13. Source maps in (Figure 4(a)) exhibit jet-structure that agrees with the flux variation. The total flux-density reaches the highest values around MJD 55300, then a rapid decrease in the flux-density is observed. Again, the flux increases and another peak is observed at MJD 56000. After that, another decrease occurs where the source-maps agree with the variability of the flux-density. The maps show jet-structures at MJD 56203 and MJD 56266 (Figure 4(b)). At MJD 57700, we notice a further decrease, which is consistent with the source-maps with jet-structures (Figure 4(c)). We can detect source structure variability by looking at short-term flux variations. We also note that this opposes the flux variability index that was determined over the whole light curve series.

In the S-band light curve (Figure 5), variations in total flux density are observed. It can be seen that the S-band light curve lags the X-band light curve by about one year. Between MJD 52300 to MJD 53000, the flux density falls down which is juxtaposed with the source-maps. At first, jet-structure appears, which soon disappears and results in apoint-like structure at MJD 52479 (Figure 5(a)). After MJD 52500, the flux density again increases and jet-structure reappears (Figure 5(b)). From MJD 57600 to MJD 58000, the flux density increases. The source-maps of 2017 show that the source exhibits jet structures throughout the year (Figure 5(c)). The light-curve shows that the flux-density decreases at MJD 58731 and after that it increases. In 2020, the source-maps show a point-like structure and then in 2021, jet-structures. In March 2021, the source-map again shows a point-like structure.

### 3.3 Source Structure Analysis

After analyzing flux variability index and light curve analysis, we present structure analysis of each source using available contour maps. Multi-epoch VLBI source maps from RFC database have been considered for this purpose.

#### 3.3.1 0048-427 (J0051-4226)

The source 0048-427 has 19 VLBI maps available between 2002 and 2020. The mean fluxes of the source at S/X-band are 0.52 Jy and 0.65 Jy respectively, and the flux variability indices are 0.34 and 0.55 respectively. The S-band light curve of this source shows no rapid variations throughout the epochs, the S-band fluxes have a standard deviation of 0.18 Jy. The X-band light curve also shows no variation until MJD 58000.

We have nine VLBI maps of the source in 2017. Available source maps indicate jet-structure variability in the source. The X-band maps indicate the appearance and disappearance of jet-structure over a period of three months. However, at S-band, the source appears to be compact in nature in the majority of epochs.

#### 3.3.2 0104-408 (J0106-4034)

The source 0104-408 has 68 VLBI source-maps available from 1994 to 2018. Mean fluxes are 0.92 Jy and 1.88 Jy and the flux variability indices are 0.4 and 0.44 at S/X-bands. The light curves of the source for both S and X bands show similar variations. But a time delay is observed between the light-curves, the X-band light-curve is ahead of the S-band.

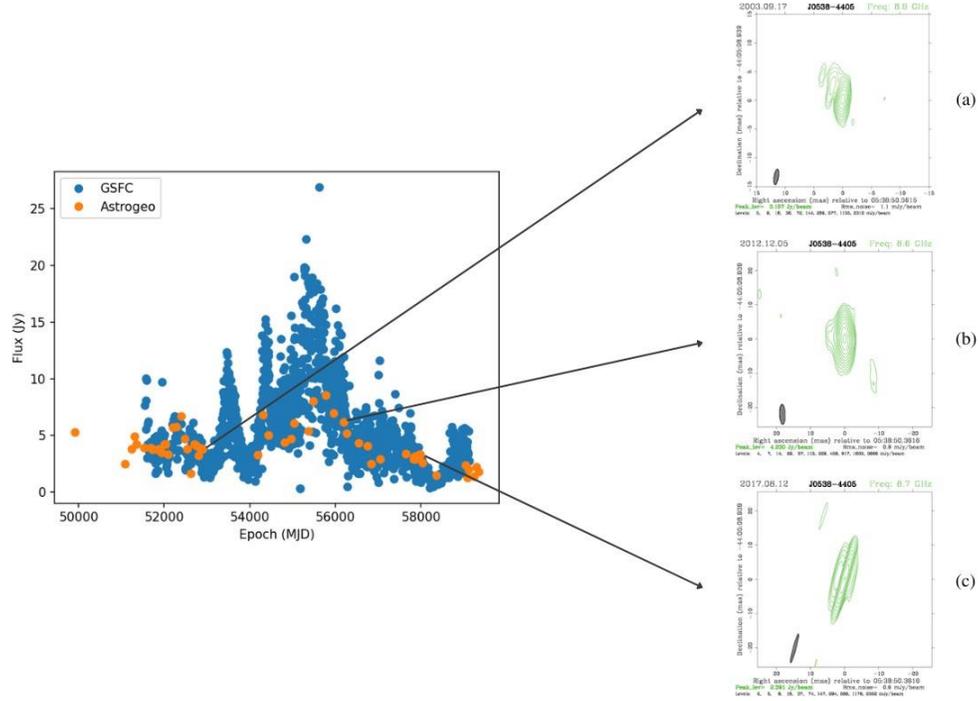

**Figure 4:** Light curve of the source 0537-441 at X-band. Orange dots show fluxes of VLBI images from RFC, and blue points are VLBI flux data from GSFC. (a) Jet-structure appeared at MJD 52899, (b) Jet-structure at MJD 56266 at X-band, (c) Jet-structure appeared at MJD 57977. All the VLBI maps are available on rfc_2022c.

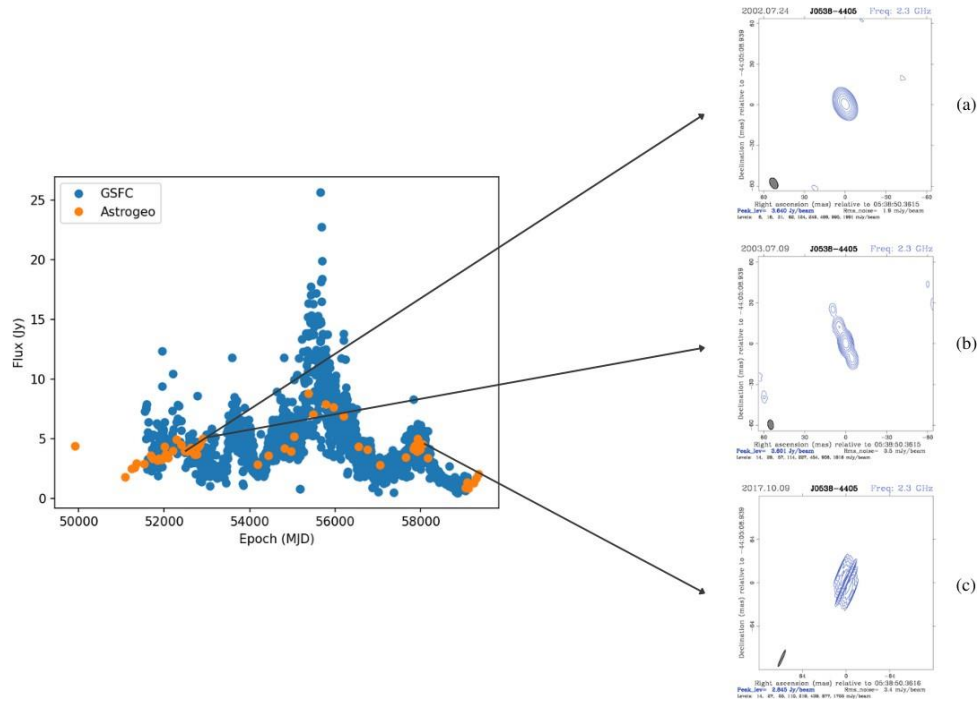

**Figure 5:** Light curve of the source 0537-441 at S-band. Orange dots show fluxes of VLBI images from RFC, and blue points are VLBI flux data from GSFC. (a) Compact structure appeared at MJD 52479, (b) Jet-structure at MJD 52889, (c) Jet-structure at MJD 58006. All the VLBI maps are available on rfc_2022c.

|           |           |       |        | X-band data |         |       | S-band data |         |       |
|-----------|-----------|-------|--------|-------------|---------|-------|-------------|---------|-------|
| B1950 Name | J2000 Name | z    | Epochs | Mean        | Std Dev | $F_I$ | Mean        | Std Dev | $F_I$ |
| 0048-427  | J0051-4226 | 1.749 | 22     | 0.658       | 0.366   | 0.556 | 0.528       | 0.182   | 0.345 |
| 0104-408  | J0106-4034 | 0.584 | 68     | 1.883       | 0.838   | 0.445 | 0.923       | 0.376   | 0.407 |
| 0332-403  | J0334-4008 | 1.445 | 28     | 1.638       | 0.601   | 0.366 | 1.548       | 0.404   | 0.261 |
| 0537-441  | J0538-4405 | 0.894 | 59     | 5.358       | 3.482   | 0.649 | 4.446       | 2.774   | 0.623 |
| 1104-445  | J1107-4449 | 1.598 | 15     | 2.126       | 0.932   | 0.438 | 1.466       | 0.40    | 0.272 |
| 1349-439  | J1352-4412 | 0.050 | 10     | 0.254       | 0.088   | 0.345 | 0.208       | 0.064   | 0.308 |
| 1424-418  | J1427-4206 | 1.522 | 52     | 4.775       | 3.341   | 0.699 | 2.609       | 1.267   | 0.485 |
| 1451-400  | J1454-4012 | 1.810 | 30     | 0.289       | 0.113   | 0.392 | 0.374       | 0.164   | 0.439 |
| 2052-474  | J2056-4714 | 1.489 | 28     | 3.112       | 2.271   | 0.729 | 1.607       | 0.586   | 0.364 |
| 2106-413  | J2109-4110 | 1.406 | 18     | 0.395       | 0.317   | 0.803 | 0.591       | 0.314   | 0.532 |
| 2333-415  | J2336-4115 | 1.058 | 13     | 0.305       | 0.125   | 0.408 | 0.256       | 0.074   | 0.290 |

**Table 2:** Col 1 and Col 2 represent the B1950 and J2000 names of the sources. Col 3 denotes the redshift of the sources. Col 4 shows the number of epochs in which observations were done. Col 5,6 and 7 represent the mean flux, standard deviation of fluxes and flux variability index respectively of the respective sources at X-band. Col 8,9 and 10 represent the mean flux, standard deviation of fluxes and flux variability index respectively of the respective sources at S-band.

The VLBI maps over all the epochs indicate the appearance and disappearance of jet-structure over a period of four months. In between 2007 and 2012, the jet structure appears every three to six months but in the year 2017, the change in the jet-structure is more frequent about 15 days. Both the S/X-band images show the same behaviour which agrees with the light curves.

### 3.3.3 0332-403 (J0334-4008)

The source 0332-403 has mean fluxes and flux variability indices of 1.54 Jy and 0.26 at S-band and 1.63 Jy and 0.36 at X-band. It has 28 VLBI maps between 2005 and 2021. In 2017, the source has 15 images in a row which evince jet-structures.

From the light curves, it is seen that the source has large deviations in its mean flux. At the same time, a different scenario is noticed at S/X-bands. It is interesting that at MJD 55000, the total flux-density at X-band decreases and the corresponding flux-density at the S-band increases. However, between MJD 55600 and MJD 56100, the X-band flux-density increases but the same at the S-band decreases. After MJD 57800, both the S and X-band flux-densities present the same behavior, they both start falling. Physically, this corresponds to the emergence of new jet-components. This agrees with the source maps of the corresponding epochs, they show jet-structures at both S/X-bands. After MJD 58400, the X-band flux-density rises more rapidly in comparison to the S-band flux-density.

### 3.3.4 0537-441 (J0538-4405)

For this source, there are 56 VLBI contour maps between 1995 and 2021. We have calculated mean flux-densities as 4.44 Jy and 5.35 Jy and flux variability indices as 0.62 and 0.64 at S/X-band respectively.

The analysis of the light curves at S- and X-band of this source is discussed in section 3.2. Between 1995 and 2018, jet-components can be seen on the source-maps of the source. At MJD 59037, the map shows no sign of jet-components, it shows point-like structure. The jet-structure again appears after four months at MJD 59297.

At S-band, between 1998 and 2003, the source exhibit jet-structures. At MJD 51840, it becomes point-like, this structure reappears after five months. The source-maps between MJD 52038 and MJD 52211 also show point-like structure of the source. Though we don't have enough maps in 2007 and 2008, but the point-like structures appear at MJD 54439 and MJD 54817. In 2017, the source has ten VLBI source-maps, which show that the source emits jet throughout the year. The behavior of the source in 2020 is discussed in section 3.2.

### 3.3.5 1104-445 (J1107-4449)

This source has mean flux-densities of 1.46 Jy and 2.12 Jy and flux variability indices 0.27 and 0.43 for S- and X-band respectively. The light-curves show that both the S/X flux-densities decreases first, but the rate of decreases is higher in case of X-band. At X-band, the flux-density starts rising from MJD 55666, but at S-band it starts rising from MJD 56666. Here the X-band flux density leads that of the S-band. After MJD 58000, the X-band flux-density increases with a greater first derivative.

This RFC does not have enough source maps to analyze the change in the source structure. It has two observations each year, so we cannot make any obvious conclusion amid these observations. But if we analyze the available source-maps at X-band after MJD 58000, we can find that the source exhibits jet-structures. This is in accordance to the rapid increase of flux-density in this period.

### 3.3.6 1349-439 (J1352-4412)

Between 2001 and 2021, the source 1349-439 has only nine contour maps. We have calculated the mean flux-densities as 0.20 Jy and 0.25 Jy and flux variability indices as 0.30 and 0.34 for S- and X-band respectively of the source. The variation of flux-density is much less about the mean for both S/X-bands. Both the light-curves mostly follow the same pattern. Between MJD 51786 and 54570, the flux-density decreases and then they start increasing. After MJD 57500, both the flux-densities rise, but the X-band density reaches a higher value. We don't have enough source-maps for this source too to analyze the source structure.

The radio fundamental catalog has two source-maps in 2021 with a gap of two months, which shows that the source has minimal jet-structures and at MJD 59353, the source appears to be point-like at X-band.

### 3.3.7 1424-418 (J1427-4206)

This source has mean flux-densities of 2.60 Jy and 4.77 Jy and flux variability indices 0.48 and 0.69 for S- and X-band respectively. The light curves for both S/X-bands follow the same pattern, but the flux-density peaks reach higher values at X-band in comparison to S-band. The highest peak of the flux-density at X-band has a value of 33.57 Jy, whereas at S-band it has a value of 13.77 Jy. The X-band light-curve leads the S-band. After MJD 52000, both the fluxes increase, the X-band flux-density shows a greater rate of increase. Between MJD 53621 and 54000, interestingly the X-band density drops, but that for the S-band density moves up. After MJD 56000, both the flux-densities show rapid rise, where both of them reach their highest values.

The source has 54 contour maps between 1994 and 2021. In the period from 1998 to 2003, we have seen the variation of flux-density. Consequently, in this period, the S/X source maps show jet-structures. Moreover, between MJD 54600 and 55070 where, the flux-densities increases a little and between MJD 56650 and 56850, where the flux-density rises rapidly, all the source-maps in these periods show jet-structure. Also, after MJD 59000, the maps exhibit jets and the light curves show declination in flux-densities.

### 3.3.8 1451-400 (J1454-4012)

This source has mean fluxes of 0.37 and 0.28 Jy, flux variability indices of 0.43 and 0.39 at S- and X-band respectively. This source has twenty-nine source-maps between 1999 and 2018. Between MJD 53000 and 54000, the S-band flux-density increases, whereas the X-band flux-density decreases. Thereafter, between MJD 54000 and MJD 56000, the behavior of the light-curves turns over, the S-band density falls and that of X-band rises. At MJD 58000, both S/X flux-densities increase and the corresponding source maps show jet-like structures.

Between MJD 51500 and 53000, the source-maps show jet-structures, but the source-map at MJD 52766 shows that the source exhibits point-like structure at S-band. At X-band, the source seems to be compact at MJD 51574. This source structure reappears at MJD 52290. At MJD 58000, both the S/X flux-densities rise and the correspondingsource-maps show jet-structures.

### 3.3.9 2052-474 (J2056-4714)

The source has 27 contour-maps from 1999 to 2017. We have calculated the mean flux-densities of this source as 1.60 Jy and 3.11 Jy and flux variability indices as 0.36 and 0.72 at S and X-band respectively. Similar to the source discussed in section 3.3.7 (source 1424-418), the peaks in the light-curve also have higher values at X-band than the S-band. The X-band light-curve also leads the S-band curve. Between MJD 51200 and 53000, the flux-densities fall, then they start increasing simultaneously. Both the S/X plot reach their highest value at MJD 55440.

Between 1999 and 2003, we have 17 maps which show jet-structures, except the maps at MJD 51938 and MJD 52038, where the source is compact for both S/X-band. After 2003, the radio fundamental catalog does not have enough source-maps to identify the nature of the source structure precisely. In accordance to the light-curves, between MJD 56000 and 57000, the flux-density falls. In 2013, we have source-maps at two adjacent epochs (MJD 56497 and 56546) in which jet components evolve.

### 3.3.10  2106-413 (J2109-4110)

This source has mean fluxes of 0.59 Jy and 0.39 Jy, flux variability indices 0.53 and 0.80 for S- and X-band respectively. From the light curves, it is seen that between MJD 52000 and 56000, the flux-densities decrease. This indicates that jet components may appear in the source structure, but unfortunately, we don't have enough VLBI images to substantiate. From MJD 56000 to 57000, the flux-densities rise and then after MJD 57000, they again decrease and the corresponding source-maps shows jet-structures.

### 3.3.11  2333-415 (J2336-4115)

The source 2333-415 has 13 contour-maps between 2012 and 2018. it has mean fluxes of 0.25 Jy and 0.30 Jy and flux variability indices 0.29 and 0.40 at S/X-band respectively. The S-band light-curve shows that the flux-density increases up to MJD 57000 and then decrease. At X-band, however, the flux-density increases after MJD 57000. Boththe S/X-band source-maps appear to be jet-structure.

## 4  Summary and Conclusion

We presented radio source structure analysis for 11 calibrator sources at the S- and X-band. All the selected sources are below ‑40° declination and were observed in multi-epoch VLBI observing sessions. To analyse the source structure variability, we mainly relied on the available source maps to detect any visible changes in the structure and also on the light-curves that were constructed using the GSFC fluxes. Firstly, we carefully went through source maps to detect any kind of variability in the source structure at mas scales. Then we used the light-curve to see the fluctuations in flux. Here, the available multi-epoch fluxes in the GSFC database were used. Finally, we extracted total fluxes (the sum of all the CLEAN components in an RFC VLBI map) and plotted those along with the GSFC fluxes. This gives a clear perspective of whether the detected structure in VLBI maps (thus change in source flux) agrees with the flux variation detected in the GSFC flux-based light-curve. We also used a metric, flux variability index, to quantify the scale of the structural variability. In this epoch-based analysis of the sources at S/X-band, we found all the selected sources exhibit jet-like structures at some or all epochs. However, based on the multi-epoch VLBI source maps, light-curves, and flux variability index analysis we analyzed the magnitude of source structure that can be used to quantify whether asource is suitable as a calibrator or not. Sources with extended jet structures, random fluctuations in the light curve, and higher flux variability index (>0.3) have been considered as not suitable for calibrators and should be observed for more analysis. Therefore, at S-band, we found three sources to be suitable candidates as calibrators. The rest of the sources may be used as calibrators, but we recommend more rigorous source structure analysis of these sources. At X-band, all the sources have flux variability index greater than 0.3 which indicates that these sources have jet-likestructures consistent over all the epochs. There are six sources having flux variability index between 0.3 and 0.5. Therest of the sources have flux variability index higher than 0.5. The flux variability indices also agree with the light curve analysis and also jet-like structures in the source maps. Overall, we recommend all the selected sources to be monitoredregularly to analyze their suitability as calibrators.

## 5  Acknowledgements

## References

Beasley, A., Gordon, D., Peck, A., Petrov, L., MacMillan, D., Fomalont, E., and Ma, C. (2002). The vlba calibrator survey—vcs1. *The Astrophysical Journal Supplement Series*, 141(1):13.

Charlot, P., Jacobs, C. S., Gordon, D., Lambert, S., de Witt, A., Böhm, J., Fey, A. L., Heinkelmann, R., Skurikhina, E., Titov, O., Arias, E. F., Bolotin, S., Bourda, G., Ma, C., Malkin, Z., Nothnagel, A., Mayer, D., MacMillan, D. S., Nilsson, T., and Gaume, R. (2020). The third realization of the International Celestial Reference Frame by very long baseline interferometry. , 644:A159.

Fomalont, E., Petrov, L., MacMillan, D., Gordon, D., and Ma, C. (2003). The second vlba calibrator survey: Vcs2. *The Astronomical Journal*, 126(5):2562.

Kovalev, Y. Y., Petrov, L., Fomalont, E., and Gordon, D. (2007). The fifth vlba calibrator survey: Vcs5. *The Astronomical Journal*, 133(4):1236.

Ma, C., Arias, E., Eubanks, T., Fey, A., Gontier, A.-M., Jacobs, C., Sovers, O., Archinal, B., and Charlot, P. (1998). The international celestial reference frame as realized by very long baseline interferometry. *The Astronomical Journal*, 116(1):516.

Petrov, L. (2016). Vlba calibrator survey 9 (vcs-9). *arXiv preprint arXiv:1610.04951*.

Petrov, L., de Witt, A., Sadler, E. M., Phillips, C., and Horiuchi, S. (2019). The second lba calibrator survey of southern compact extragalactic radio sources–lcs2. *Monthly Notices of the Royal Astronomical Society*, 485(1):88–101.

Petrov, L., Kovalev, Y. Y., Fomalont, E., and Gordon, D. (2005). The third vlba calibrator survey: Vcs3. *The Astronomical Journal*, 129(2):1163.

Petrov, L., Kovalev, Y. Y., Fomalont, E., and Gordon, D. (2006). The fourth vlba calibrator survey: Vcs4. *The Astronomical Journal*, 131(3):1872.

Petrov, L., Kovalev, Y. Y., Fomalont, E., and Gordon, D. (2008). The sixth vlba calibrator survey: Vcs6. *The Astronomical Journal*, 136(2):580.

Petrov, L., Phillips, C., Bertarini, A., Murphy, T., and Sadler, E. M. (2011). The lba calibrator survey of southern compact extragalactic radio sources–lcs1. *Monthly Notices of the Royal Astronomical Society*, 414(3):2528–2539.

Shabala, S.S. et al. (2014), The effects of frequency-dependent quasar variability on the celestial reference frame. Journal of Geodesy, 88:575-586.

Wang, F., Yang, J., Fan, X., Hennawi, J. F., Barth, A. J., Banados, E., Bian, F., Boutsia, K., Connor, T., Davies, F. B., et al. (2021). A luminous quasar at redshift 7.642. *The Astrophysical Journal Letters*, 907(1):L1.